 \documentclass[final,5p,times,twocolumn,authoryear]{elsarticle}

\usepackage{amssymb}
\usepackage{lipsum}
\usepackage{hyperref}

\journal{Fundamental Plasma Physics}

\begin{document}

\begin{frontmatter}



\title{Star formation across cosmic time}


\author[first]{Jonathan Freundlich}
\affiliation[first]{organization={Université de Strasbourg, CNRS UMR 7550, Observatoire astronomique de Strasbourg},
            addressline={11 rue de l'Université}, 
            city={Strasbourg},
            postcode={67000}, 
            country={France}}
 
\begin{abstract}
The interstellar medium of galaxies is composed of multiple phases, including molecular, atomic, and ionized gas, as well as dust. Stars are formed within this medium from cold molecular gas clouds, which collapse due to their gravitational attraction. Throughout their life, stars emit strong radiation fields and stellar winds, and they can also explode as supernovae at the end of their life. These processes contribute to stirring the turbulent interstellar medium and regulate star formation by heating up, ionizing, and expelling part of the gas. However, star formation does not proceed uniformly throughout the history of the Universe and decrease by an order of magnitude in the last ten billion years. To understand this winding-down of star formation and assess possible variations in the efficiency of star formation, it is crucial to probe the molecular gas reservoirs from which stars are formed. In this article following my presentation at the 10th International Conference on Frontiers of Plasma Physics and Technology held in Kathmandu from 13-17 March 2023, I review some aspects of the multiphase interstellar medium and star formation, with an emphasis on the interplay between neutral and ionized phases, and present recent and ongoing observations of the molecular gas content in typical star-forming galaxies across cosmic time and in different environments. I also present some of our understanding of star-forming galaxies from theoretical models and simulations. 

\end{abstract}




\end{frontmatter}




\section{Introduction}
\label{introduction}

Observations in the microwave domain of the electromagnetic spectrum reveal that we are surrounded by a bath of photons emitted at a moment when the Universe was extremely hot, dense, and homogeneous. This {\it cosmic microwave background} gives us a snapshot of the Universe when it had cooled down sufficiently for protons and neutrons to combine into neutral hydrogen atoms and for photons to start propagating. This epoch is referred to as {\it recombination} (although protons and electrons were not combined before), and the hot, dense plasma that preceded it is thought to have been opaque to electromagnetic radiation since the mean free path of photons was extremely short. Structures emerge from the homogeneous primordial Universe through a competition between the gravitational attraction, which tends to enhance local over-densities by attracting the surrounding matter, and the expansion of the Universe, which tends to dilute such over-densities. 

In the current standard model of cosmology based on Einstein's general relativity theory, the structure of the Universe and its expansion is intrinsically related to its matter and energy content. The minute fluctuations of the cosmic microwave background and the systematic recession of galaxies from us imply in this context a universe constituted by 69\% of an unknown form of energy named {\it dark energy}, which drives the expansion of the Universe, 26\% of an unknown form of matter called {\it dark matter}, and 5\% of ordinary matter, often refereed to as {\it baryons} \citep{Planck2016-cosmo}. Since dark matter represents 84\% of the total matter content of the Universe, it drives the formation of structures at very large scale, forming a {\it cosmic web} of over-densities, but it remains relatively diffuse given that it does not radiate away energy. 
Baryons on their part can emit photons, lose energy, and hence condense further due to the gravitational attraction at the center of the more diffuse {\it dark matter haloes}. Post-recombination, baryons are in the form of a neutral gas mostly composed of hydrogen atoms\footnote{In addition to hydrogen, primordial nucleosynthesis also produces deuterium, helium, lithium, and isotopes.}. When this gas collapses due to its own gravitational attraction, temperature and pressure can get so high that deuterium and hydrogen fusion reactions ignite: stars are born. Throughout their life, stars emit strong radiation fields, stellar winds, and can explode as supernovae, which not only ionize their surrounding {\it interstellar medium} but also eventually most of the Universe. 

\begin{figure*}[ht]
	\centering 
	\includegraphics[width=1\textwidth, trim={0 0 0 0.5cm},clip]{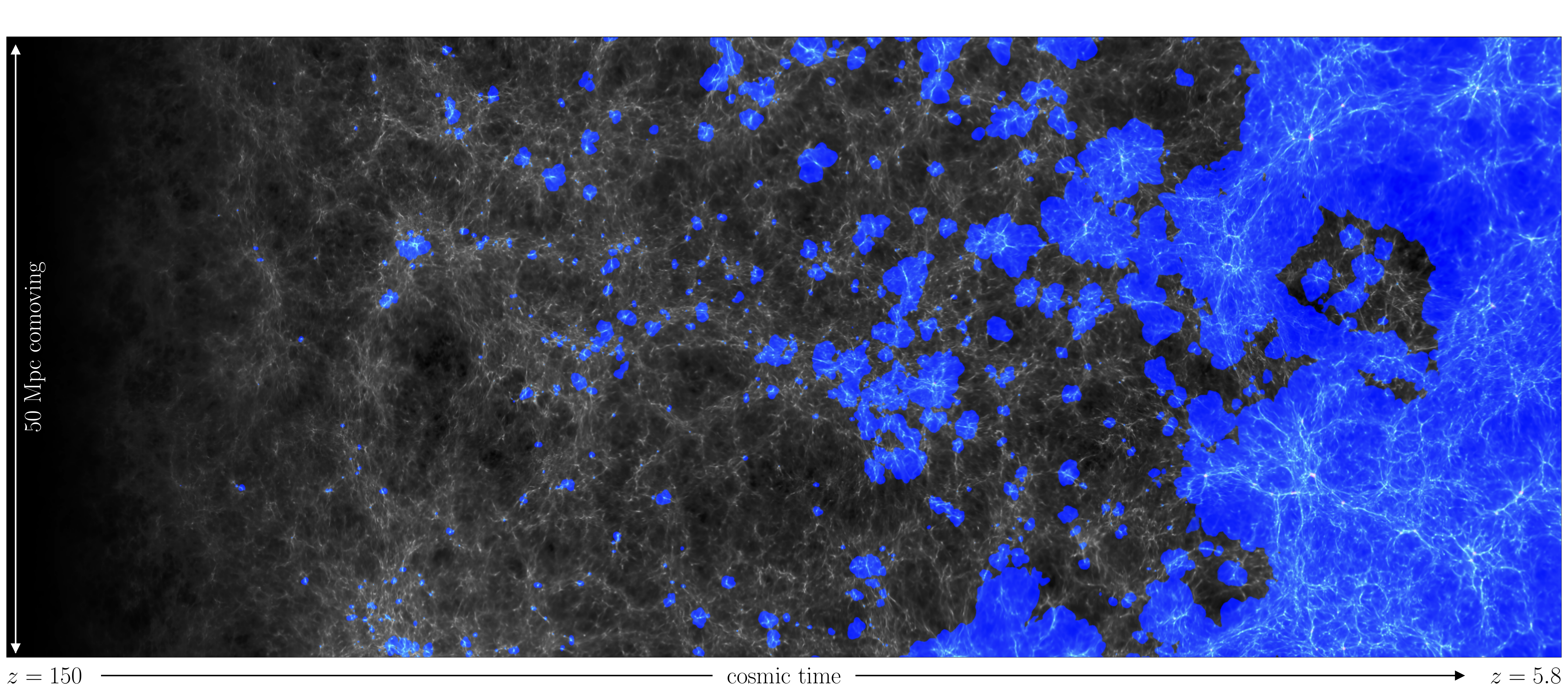}	
    \includegraphics[width=1\textwidth, trim={0 1.2cm 0 0cm},clip]{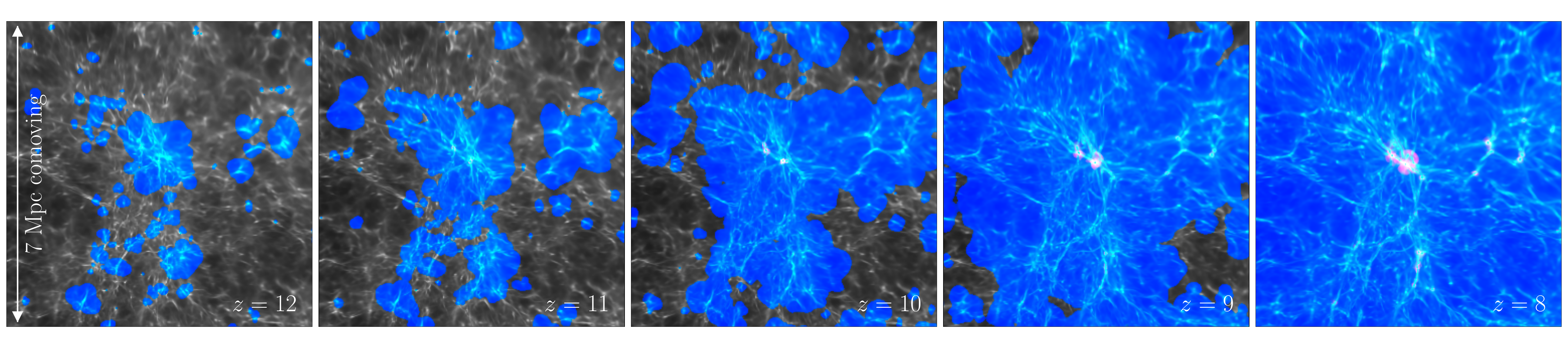}	
	\caption{Illustration of the progressive reionization of the Universe, from the CoDA II simulation by \cite{Ocvirk2020}. 
    {\it Top panel:} Time evolution of a planar slice through the simulation cube, with time increasing from left to right between redshift $z=150$ shortly after recombination to $z=5.8$, i.e. during the first Gyr of the history of the Universe. The vertical axis corresponds to 50 comoving Mpc. 
    {\it Bottom panels:} Slices zooming around a massive cluster of galaxies at different times, from $z=12$ to $z=8$. 
    Blue regions are photo-heated, small red regions are heated by supernovae feedback and accretion shocks, green regions corresponds to regions where ionization is ongoing or incomplete. Brightness indicate the gas density. The Universe globally evolves from neutral at recombination to ionized, but small pockets of neutral gas remain within galaxies, where stars are formed.}
	\label{fig_reionisation}%
\end{figure*}

Fig.~\ref{fig_reionisation} shows this {\it reionisation} of the Universe during the first billion year following recombination, as seen in a cosmological simulation by \cite{Ocvirk2020}. Ionized bubbles progressively percolate through the neutral gas until most of the volume is ionized. Small pockets of neutral gas however still remain within galaxies at the center of dark matter haloes, where new stars can continue to form. 
This large scale simulation of a portion of the Universe follows the ionized gas fraction and its effect on the temperature of the gas, which in turn affect the dynamics, but it does not take into account the specific dynamics of charged particles in the presence of electromagnetic fields nor the elaborate chemistry of the interstellar gas. Other simulations at much smaller scales within galaxies can incorporate the effect of electromagnetic fields on interstellar plasma and/or part of its chemistry  \citep[e.g.,][]{Hennebelle2008, Walch2015}, but without the cosmological nor most of the galactic context. Galaxy formation and evolution indeed represents a complex multi-scale phenomenon, encompassing more than ten orders of magnitude in length, ranging from sub-parsec scales to cosmological megaparsec scales\footnote{1 parsec (pc) $= 3.09 \times 10^{16} ~\rm m$ = 3.26 ~light-years.}. Moreover, the large-scale environment can affect the interstellar medium locally, while local processes such as star formation, stellar evolution and supernova explosions can reverberate at large distances from the galaxy. This makes it challenging if not impossible to include all scales and processes simultaneously in numerical simulations. 

Despite representing only 15\% of the total matter content, baryons make up the visible Universe. Star formation is the essential step through which galaxies come into existence and from which most chemical elements that constitute us result. Furthermore, stars shape their surrounding interstellar medium, their host galaxies, the circumgalactic medium around galaxies, and leave their imprint on structures at cosmological scales. How do stars form? What is their effect on their surroundings? How did star formation proceed throughout the history of the Universe? Did star formation processes evolve across cosmic time? 
As we will see, understanding gas in galaxies, its different phases, and its interplay with stars is crucial to address these questions.
Not only because stars are formed from cold molecular gas, but also because stellar evolution, supernova explosions, and active galactic nuclei affect the surrounding gas and regulate star formation.
In the following, Section~\ref{sect_ism} focuses on star formation in the interstellar medium; Section~\ref{sect_feedback} on feedback processes resulting from stellar evolution and active galactic nuclei; Section~\ref{sect_sfh} on the cosmic star formation history; Section~\ref{sect_efficiency} on characterizing star formation processes at different epochs; and Section~\ref{sect_conclusion } concludes by presenting some perspectives.

\section{Star formation in the interstellar medium}
\label{sect_ism}

\begin{figure*}
	\centering 
	\includegraphics[width=0.48\textwidth, trim={3.8cm 2.8cm 5.7cm 11.cm},clip]{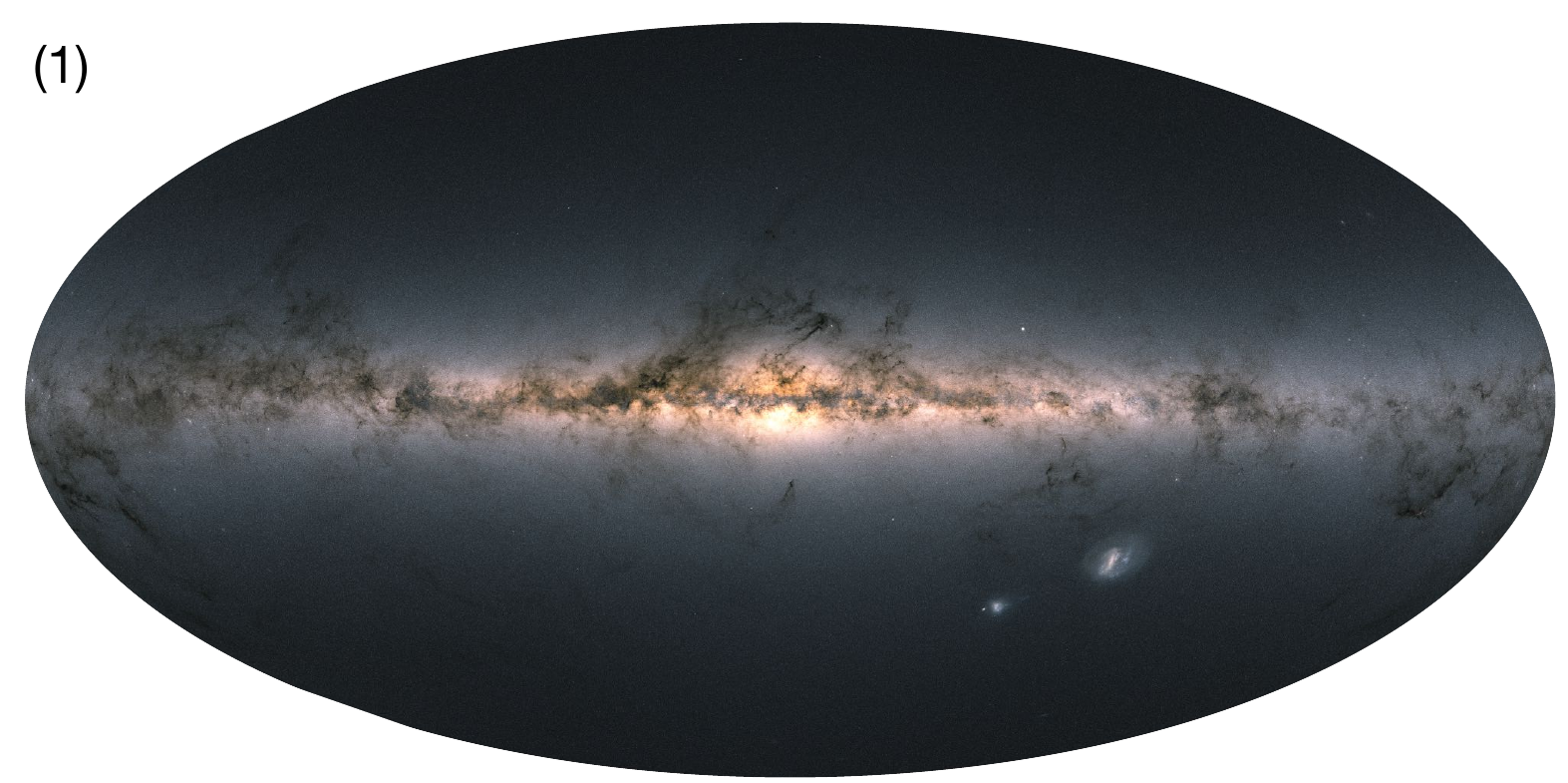}\hfill
        \includegraphics[width=0.48\textwidth, trim={3.8cm 2.8cm 5.7cm 11.cm},clip]{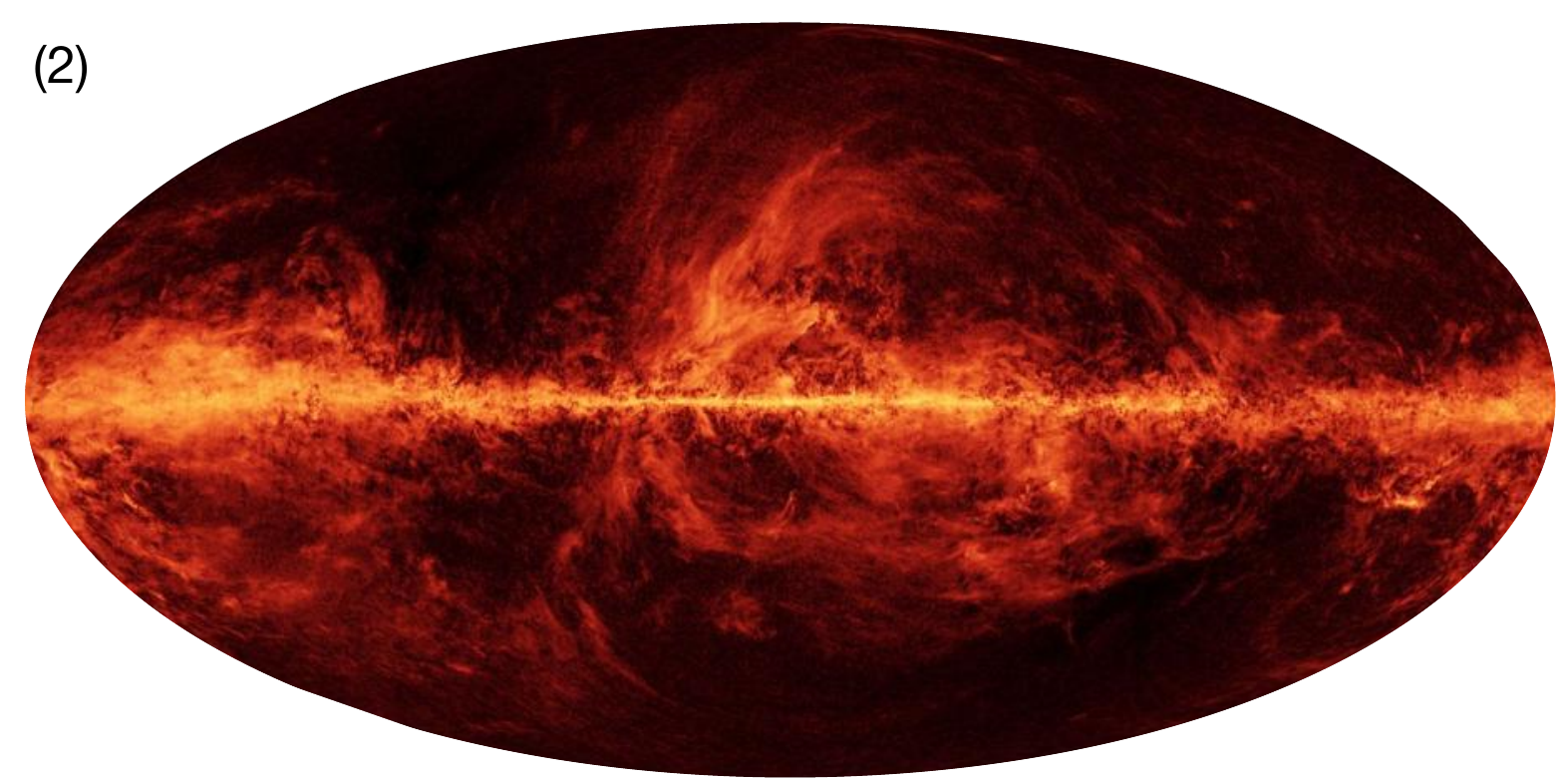}\\
        \includegraphics[width=0.48\textwidth, trim={3.8cm 2.8cm 5.7cm 11.cm},clip]{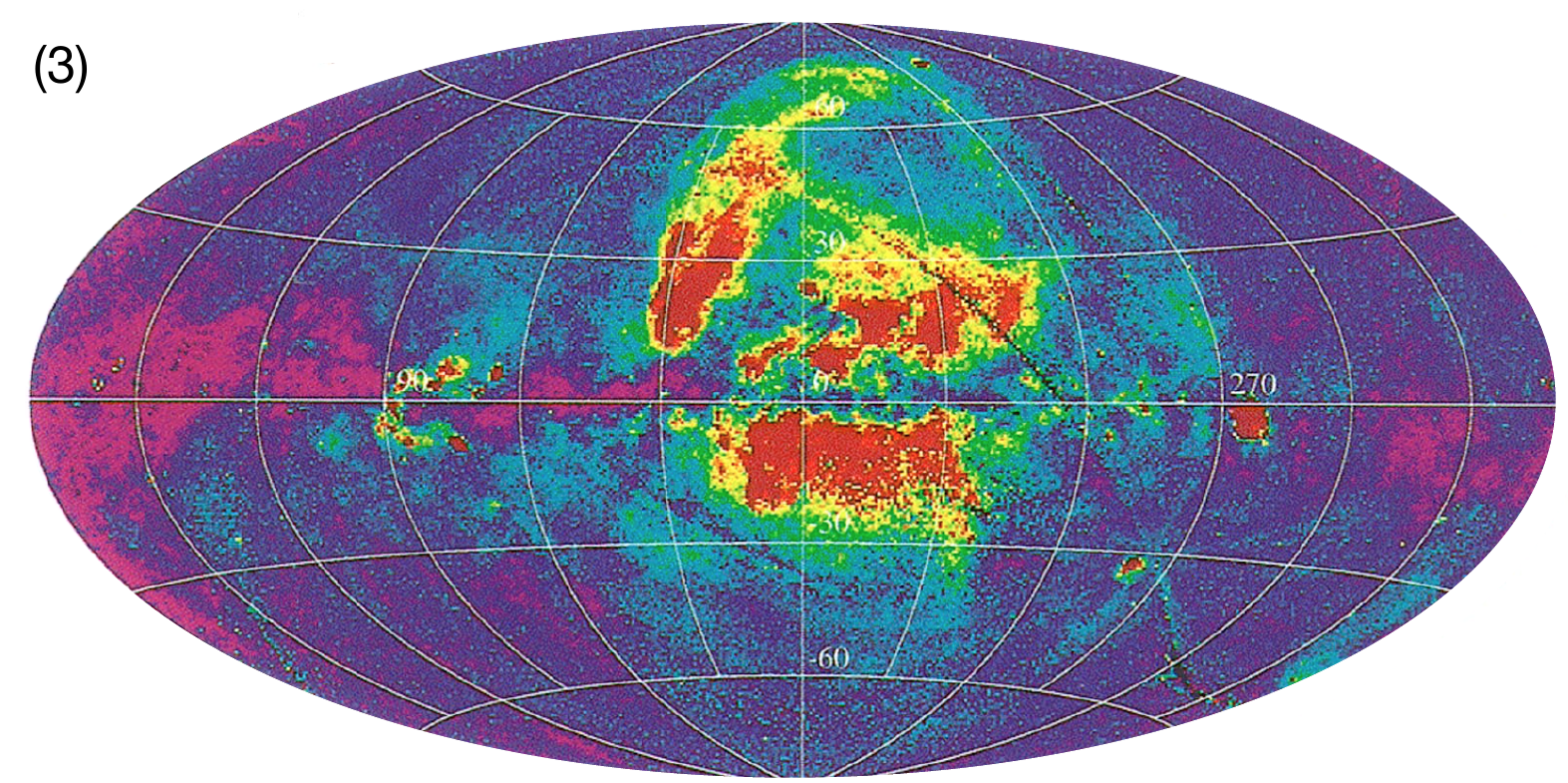}\hfill
        \includegraphics[width=0.48\textwidth, trim={3.8cm 2.8cm 5.7cm 11.cm},clip]{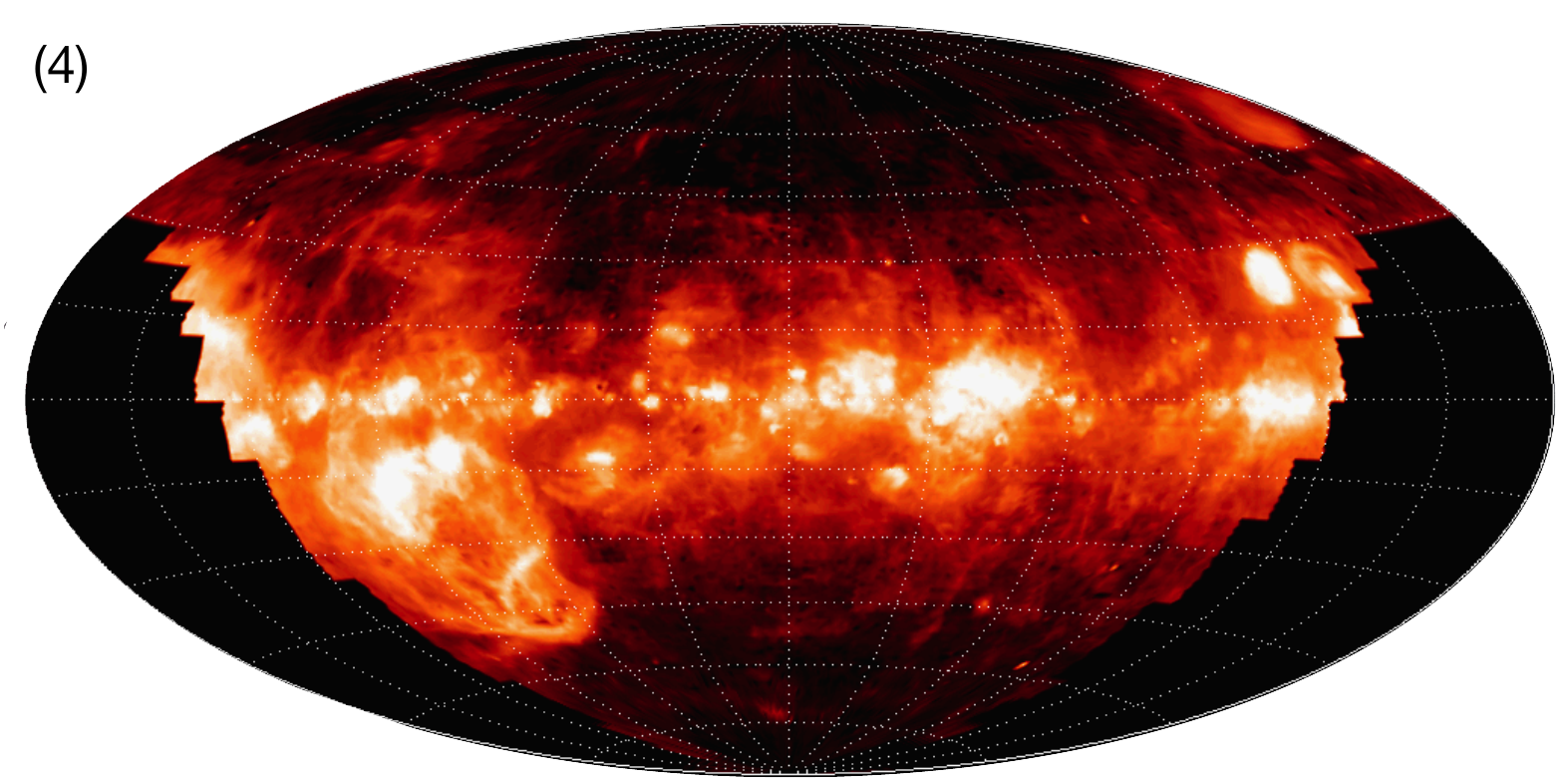}\\
        \includegraphics[width=0.48\textwidth, trim={3.8cm 2.8cm 5.7cm 11.cm},clip]{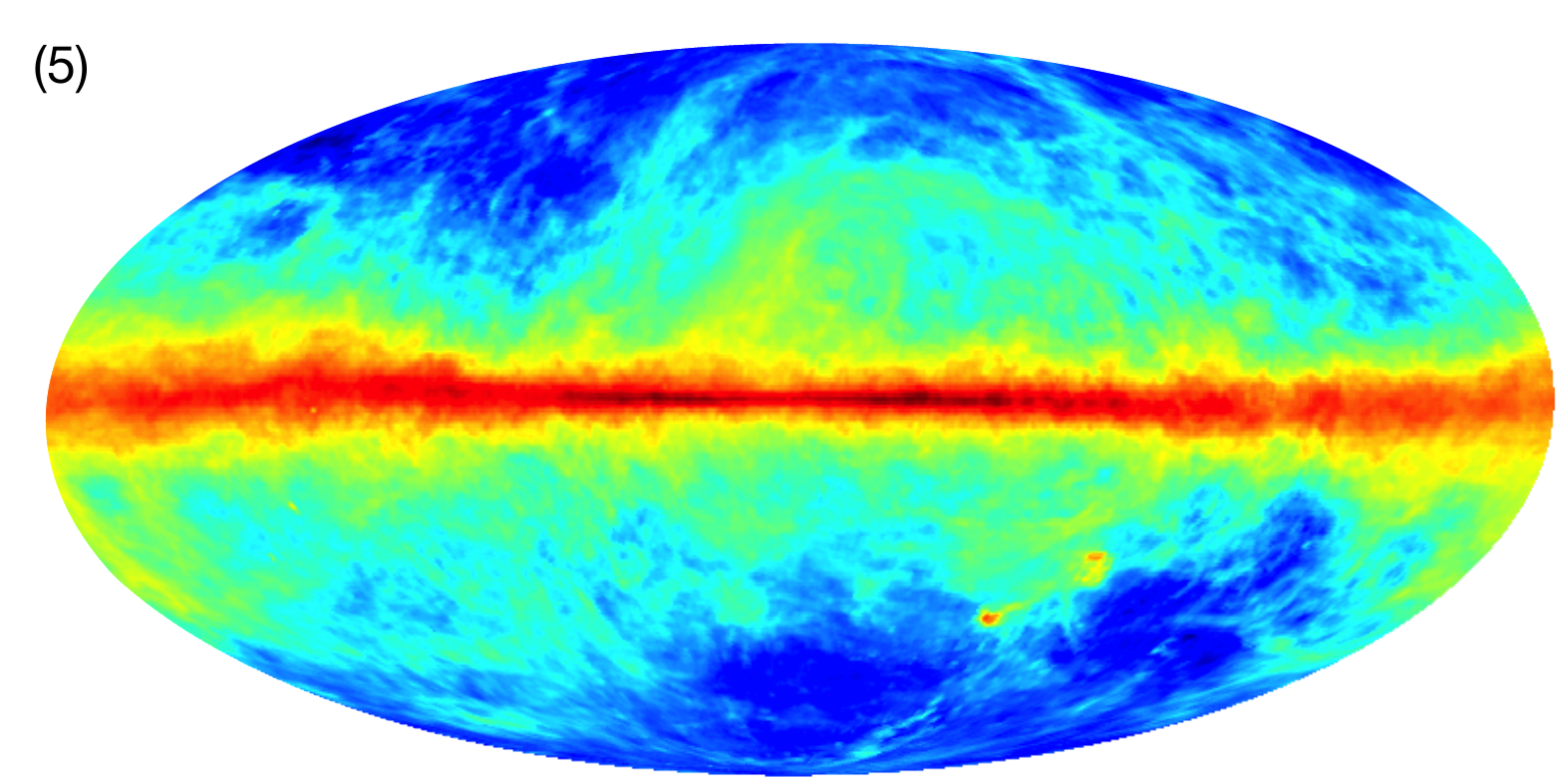}\hfill
        \includegraphics[width=0.48\textwidth, trim={3.8cm 2.8cm 5.7cm 11.cm},clip]{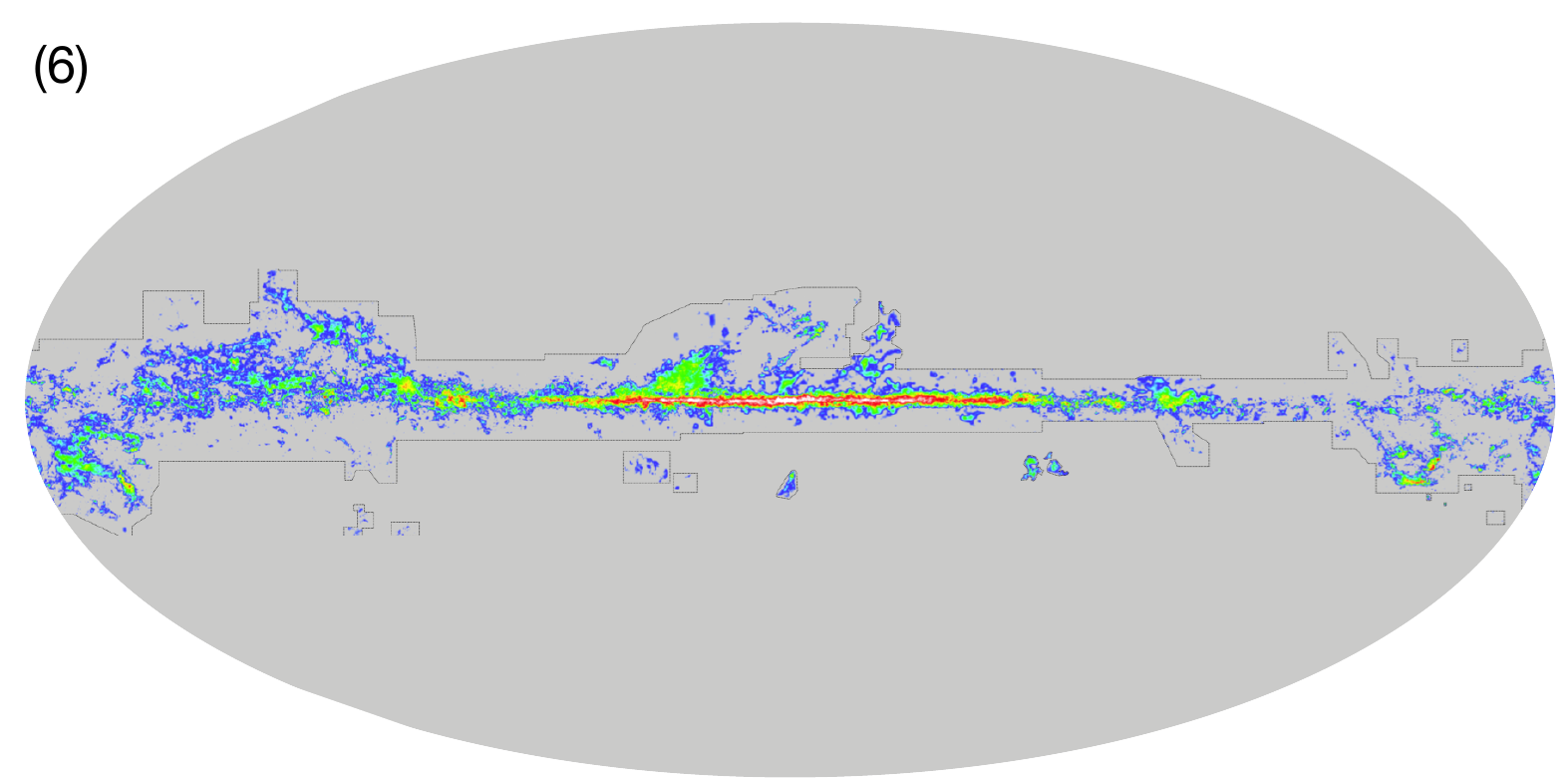}
	\caption{The Milky Way seen at different wavelength ranges in the electromagnetic spectrum, each of these ranges tracing a different component of the galaxy: 
        (1) optical light traces stars, which are obscured by dust clouds (image obtained by the Gaia space observatory; credits: ESA/Gaia/DPAC); 
        (2) the infrared continuum traces the dust itself (Planck space observatory; ESA/NASA/JPL-Caltech);
        (3) X-rays trace the hot ionized medium \citep[ROSAT space telescope; ][]{Snowden1997}; 
        (4) the H$\alpha$ infrared emission line traces ionized H\,{\sc II} regions \citep[Kitt Peak National Observatory; ][]{Haffner2003}; 
        (5) the 21cm emission line traces neutral hydrogen \citep[Dwingeloo \& Buenos Aires radio observatories; ][]{Kalberla2005}; 
        and (6) the carbon monoxide $\rm CO$ emission line traces the dense molecular gas \citep{Dame2001}. 
        } 
	\label{fig_mw}%
\end{figure*}

\begin{figure*}
	\centering 
	\includegraphics[width=1\textwidth, trim={0 20.2cm 0 0},clip]{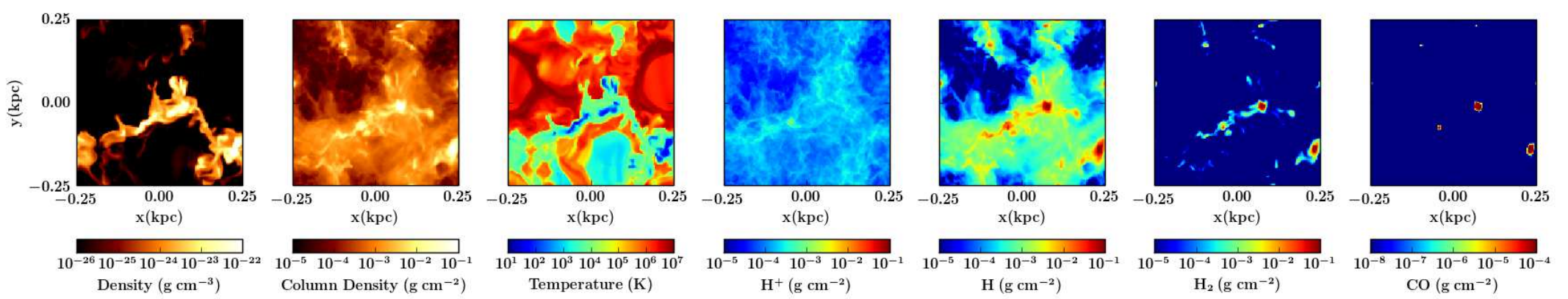}	
	\caption{Gas density slice, column density projection, gas temperature slice, and column density of ionized $\rm H^+$, atomic $\rm H$, and molecular $\rm H_2$ and $\rm CO$ at a given snapshot in a simulation of the multi-phase interstellar medium by \cite{Walch2015}. The simulation shown here includes gravity, magnetic fields, heating and radiative cooling, chemistry of $\rm H_2$ and $\rm CO$ as well as supernova explosions. The simulation focuses on a small region of a galactic disk, 500 pc in width. } 
	\label{fig_silcc}%
\end{figure*}

\begin{figure}
	\centering 
        \includegraphics[width=0.48\textwidth, trim={0.7cm 2.8cm 0.6cm 0.9cm},clip]{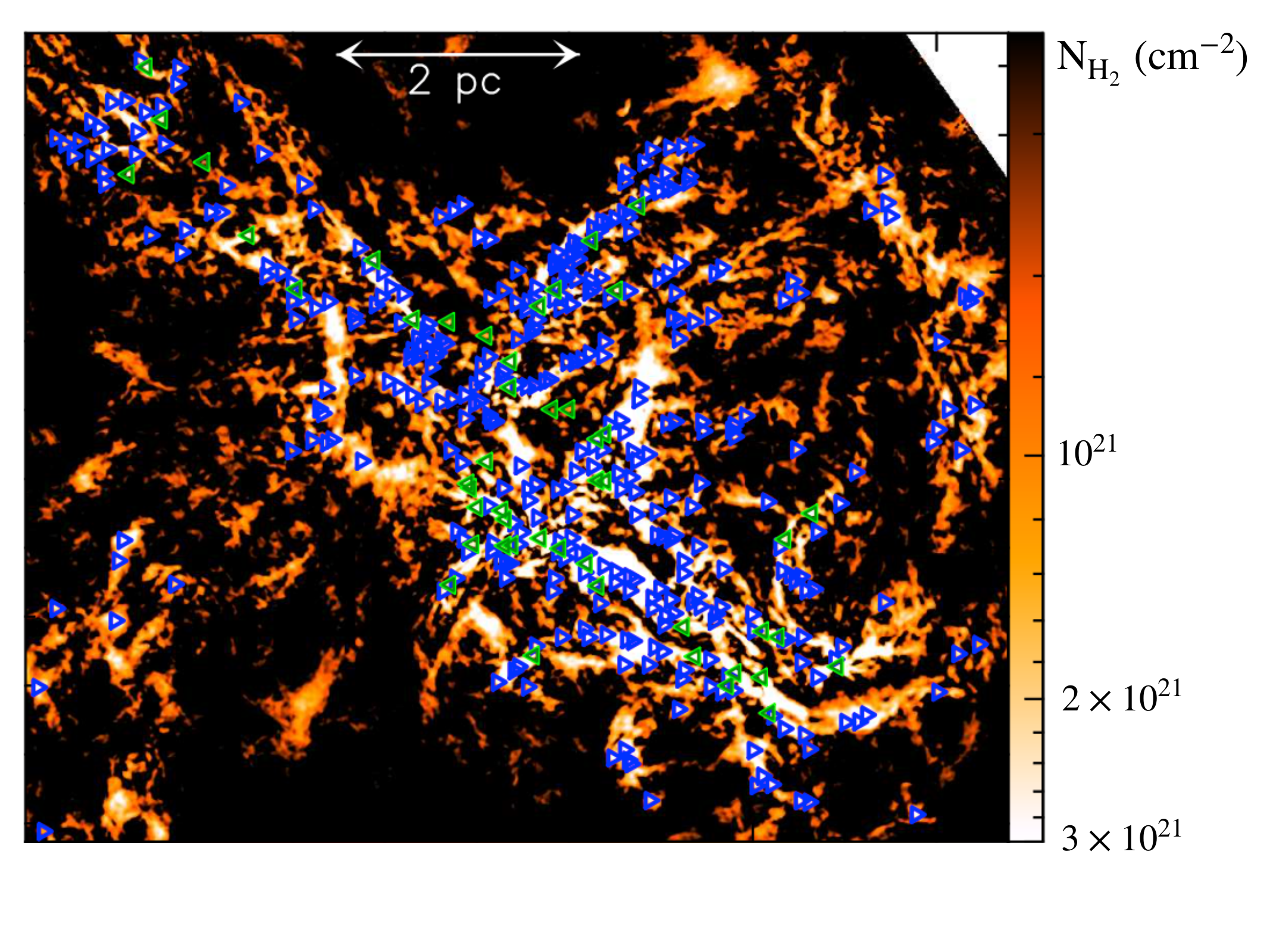}
	\caption{Column density map of a subregion of the Aquila star-forming complex of the Milky Way, highlighting the link between the filamentary over-densities and the location of pre-stellar and proto-stellar clores (blue and green triangles, respectively). The majority of pre-stellar cores (75\%) lie within the 0.1 pc-wide filaments. Figure from \cite{Konyves2015}, based on observations with the Herschel Space Observatory. 
        } 
	\label{fig_konyves}
\end{figure}

A galaxy like the Milky Way can be defined as a gravitationnally bound system that contains many stars, but it is not limited to its stars and the dark matter that surrounds it. The {interstellar medium} of a galaxy, which constitutes the rest of the matter, represents about 10\% of the visible mass. Most of it is in the form of gas, representing about 99\% of its mass, but it also contains some {\it dust}, namely small solid particles mixed with the interstellar gas, {\it cosmic rays}, namely ions and electrons whose velocities are much higher than thermal velocities, and electromagnetic {\it radiation} from stars and other sources. The interstellar medium of a galaxy hosts a variety of chemical elements, from hydrogen, helium, carbon, nitrogen, and oxygen to heavier elements. Most of the stars, gas, and dust of a galaxy like the Milky Way are located in and around a relatively thin disk whose thickness is of a few hundred pc. 
As reviewed by \cite{Draine2011}, gas in the interstellar medium can be separated into different phases: 
\begin{itemize}
    \item The {\it hot ionized medium}, or coronal gas, has a very low density ($n<0.01 ~\rm atoms/cm^3$) but fills a large fraction of the volume in and around the galactic disk. It has been heated to very high temperatures ($T\gtrsim 3\times 10^5~\rm K$) by shocks, e.g. from the blastwaves of supernova explosions, and it is collisionally ionized to high ionized states, with ions such as O\,{\sc vi} ($\rm O^{5+}$), N\,{\sc v} ($\rm N^{4+}$), or C\,{\sc iv} ($\rm C^{3+}$). 
    
    \item The {\it H\,{\sc ii} ionized gas}, where hydrogen has notably been photo-ionized by ultraviolet photons from hot stars, has densities in the range $n\sim 0.3 - 10^4 \rm ~atoms/cm^3$, from dense clouds surrounding young stars (referred to as {\it H\,{\sc ii} regions}) to the lower density intercloud medium ({\it diffuse H\,{\sc ii}} or {\it warm ionized medium}) and supernova remnants ({\it planetery nebulae}). It fills about 10\% of the volume of the disk and its temperature is of the order of $T\sim 10^4 ~\rm K$. 

    \item The {\it warm neutral medium}, or warm H\,{\sc i}, is predominantly atomic and fills about 40\% of the volume of the disk. It has a density $n\sim 0.6 \rm ~atoms/cm^3$ and a temperature $T\sim 5,000~ \rm K$.

    \item The {\it cool neutral medium}, or cool H\,{\sc i}, is also predominantly atomic. Its has a density $n\sim 30 ~\rm atoms/cm^3$, a temperature $T\sim 100~\rm K$, and fills about 1\% of the volume of the disk. 

    \item The {\it diffuse molecular gas} is similar to the cool neutral medium but with densities sufficiently large to enable the presence of $\rm H_2$ molecules. Namely, it corresponds to $n\sim 100 \rm ~atoms/cm^3$, $T\sim 50~\rm K$, and fills about 0.1\% of the volume of the disk. 

    \item The {\it dense molecular gas} is constituted by gravitationally-bound clouds where the density has reached $10^3$ to $10^6$ $\rm atoms/cm^3$ and temperatures are in the range $10-50$ K. This is where star formation occurs. 

\end{itemize}
In addition to these different phases, stars can generate outflows through which they lose part of their mass. These winds blow in the interstellar medium with densities in the range $1-10^6\rm~atoms/cm^3$, temperatures between $50-10^3~\rm K$, and velocities from a few tens to hundreds of km/s. 

\begin{figure}
	\centering 
        \includegraphics[width=0.48\textwidth, trim={0cm 16.5cm 0cm 6cm},clip]{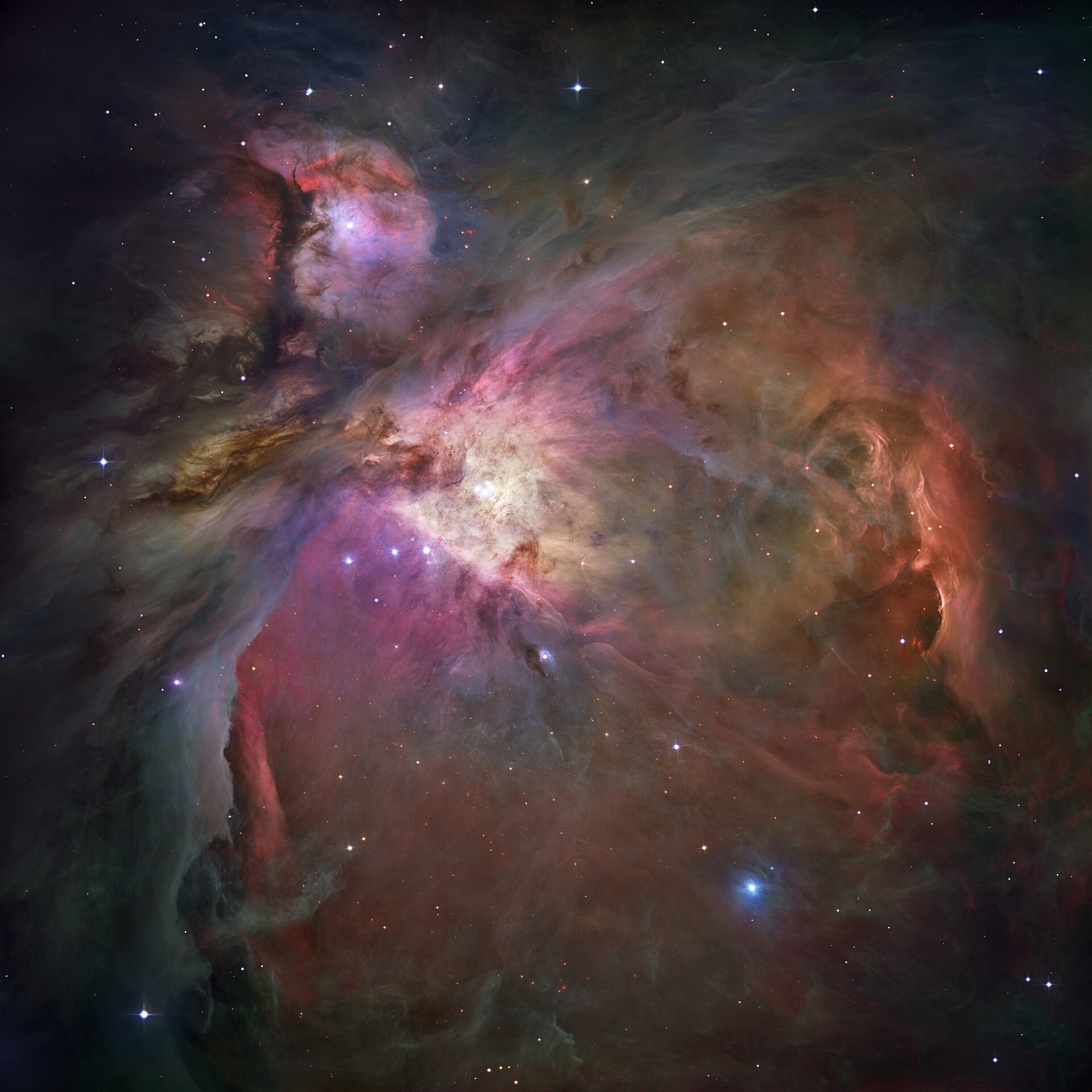}
	\caption{The Orion Nebula, a stellar nursery. Young stars at the heart of the nebula emit energetic ultraviolet ionizing radiation and winds that carve a cavity around them. Their ultraviolet light reflected at the edge of the cavity can be seen in blue; the surrounding gas they have ionized, traced by the H$\alpha$ emission line of ionized hydrogen, is shown in red. 
        Credits: NASA, ESA, M. Robberto (Space Telescope Science Institute/ESA) and the Hubble Space Telescope Orion Treasury Project Team.  
        } 
	\label{fig_orion}
\end{figure}

\begin{figure*}
	\centering 
        \includegraphics[width=1.\textwidth, trim={0cm 18.1cm 0cm 0cm},clip]{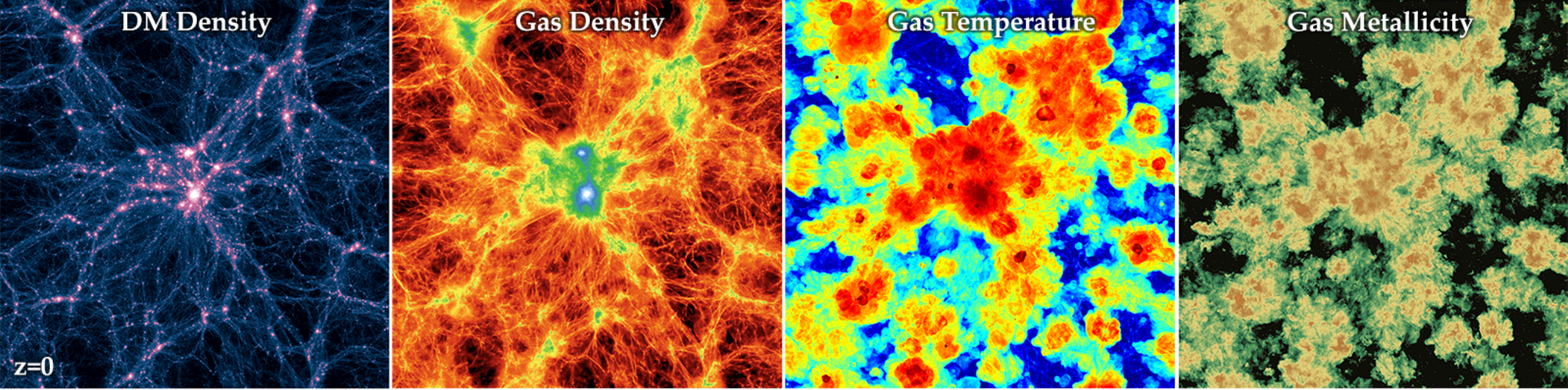}
	\caption{Dark matter density, gas density, gas temperature, and gas metallicity at redshift $z=0$ in the Illustris cosmological simulation \citep{Nelson2015}. The simulated portion of the Universe is 106.5 Mpc wide. The dark matter density traces the cosmic web, cosmic filaments, dark matter haloes at the nodes of these filaments, and a massive central cluster. The gas density follows the dark matter density at large scale. Feedback processes in galaxies result in blastwaves propagating in the intergalactic medium, which both heat up the gas and disseminate elements heavier than helium, as shown by the gas temperature and metalicity. Credits: Illustris Collaboration. 
        } 
	\label{fig_illustris}
\end{figure*}

Fig.~\ref{fig_mw} shows images of the Milky Way in different wavelength ranges of the electromagnetic spectrum, each tracing a different component of the galaxy: stars and dust, hot ionized medium,  H\,{\sc ii} regions, neutral  H\,{\sc i} gas, and molecular gas. The latter can notably be traced at radio wavelengths by the line emission of carbon monoxide (CO), which is the second most abundant molecule after $\rm H_2$ but more easily observed. Since it can be rotationally excited by collisions with $\rm H_2$ molecules in the cold interstellar medium, it provides a tracer of molecular hydrogen \citep[e.g., ][]{Combes1991, Bolatto2013}. The conversion between CO luminosity and molecular gas mass however hinges on the CO excitation, the {\it metallicity} of the gas, i.e. the amount of elements heavier than hydrogen and helium, and more generally the physical conditions of the interstellar medium. 
Fig.~\ref{fig_silcc} shows a simulated view of the interstellar medium by \cite{Walch2015}, highlighting its complex multi-phase nature. This simulation does not include all the chemical elements and reactions that take place in the interstellar medium, but it does track the ionization fraction of the gas, the formation and destruction of $\rm H_2$ and CO molecules, and the effect of supernova blastwaves. 

Stars form in the densest regions of {\it giant molecular clouds}, where gravity pulls gas particles together to form high-density cores which will later turn into stars. Giant molecular clouds are not smooth and instead exhibit intricate structures, hosting complex networks of filamentary over-densities \citep{Arzoumanian2011, Andre2014}. Such filaments can arise from intersecting shock waves in the turbulent, magnetised interstellar medium, become self-gravitating, and for the densest of them fragment into pre-stellar cores as a result of gravitational instability \citep{Inutsuka1992, Inutsuka1997, Hennebelle2013, Freundlich2014}. As shown in Fig.~\ref{fig_konyves}, from \cite{Konyves2015}, a large fraction of the gravitationally-bound pre-stellar and proto-stellar cores that may eventually or already have formed stars indeed lie within these filamentary structures. When a pre-stellar core is dense enough, deuterium fusion reactions ignite, followed by hydrogen fusion reactions. Hydrogen fusion is the main process that makes stars like our Sun shine. 
Young stars emit highly energetic ultraviolet light and generate winds that ionize and blow away the surrounding gas. This can notably be seen in Fig.~\ref{fig_orion} for the Orion Nebula, where the winds generated by the bright young stars at the center of the stellar nursery have carved a huge oval cavity.

\section{Feedback processes}
\label{sect_feedback}

Throughout their life, stars emit strong radiation fields, neutrinos, and winds. Furthermore, massive stars can also eventually explode as supernovae, directly injecting mass, momentum, and kinetic energy into the interstellar medium. Active galactic nuclei (AGN) due to the accretion around the central supermassive black holes of galaxies also induce similar effects. These different processes, collectively named {\it feedback}, can not only hinder subsequent star formation by heating up the gas, expelling part of it through powerful outflows, decreasing or suppressing the efficiency of star formation, or preventing further accretion onto galaxies by heating up the {\it circumgalactic medium}, but they also drive supersonic turbulence in the interstellar medium and disseminate elements heavier than helium formed within stars or during supernova explosions. They can at times also have a positive effect on star formation, notably by enhancing cooling with the injection of heavy elements, by catalysing H$_2$ formation with the release of free electrons during the ionization of hydrogen and helium, or by locally compressing the gas during shocks (e.g., \citealp{Loeb2013}, p. 200, \citealp{Salome2015}, \citealp{Cresci2015}).
Feedback processes can affect galaxies at larger scales, their surrounding circumgalactic medium, and possibly even the distribution of dark matter \citep[e.g., ][for the latter]{ElZant2016, Freundlich2020a}. 
Because of feedback processes, galaxies at the low and high mass ends form comparatively less stars than expected from the distribution of dark matter haloes \citep[e.g.][]{Dekel1986, Silk2012, Behroozi2013, Behroozi2019}: at low mass because the relatively shallow gravitational potentials enable strong outflows resulting from stellar evolution, at high mass because of the presence of supermassive black holes and active galactic nuclei around them. 
Observations reveal ionized gas winds driven by star formation with velocities of the order of $\sim 500\rm~km/s$ and mass outflow rates comparable to the star formation rate, while active galactic nuclei can launch winds with faster velocities $\sim 1000 -2000\rm ~km/s$ \citep[e.g.][]{Forster2019}. 
Active galactic nuclei can further launch highly collimated relativistic jets that can propagate up to Mpc-scales \citep[e.g.][]{Blandford2019}.

Fig.~\ref{fig_illustris} highlights how feedback processes, which originate from relatively localised regions around stars and active galactic nuclei, affect galaxies at large scale and their circumgalactic medium. It shows the dark matter and gas densities, the gas temperature, and the gas metallicity for the final $z=0$ snapshot of the Illustris cosmological simulation \citep{Nelson2015}, which notably includes gas cooling, star formation, stellar evolution, supernova explosions and winds, as well as the presence of supermassive black holes and feedback from active galactic nuclei. The different feedback processes result in the propagation of blastwaves in the intergalactic medium, heating the gas and disseminating heavy elements at very large scales. 
Earlier in the history of the Universe, similar processes contributed to reionize the Universe. 
If such cosmological simulations enable to reproduce the observed large scale structure of the Universe and a number of galaxy properties, they rely on sub-grid recipes to treat processes such as star formation and feedback, which physically occur beyond the resolution limit.

\begin{figure}
	\centering 
        \includegraphics[width=0.426\textwidth, trim={0cm 0cm 0cm 0cm},clip]{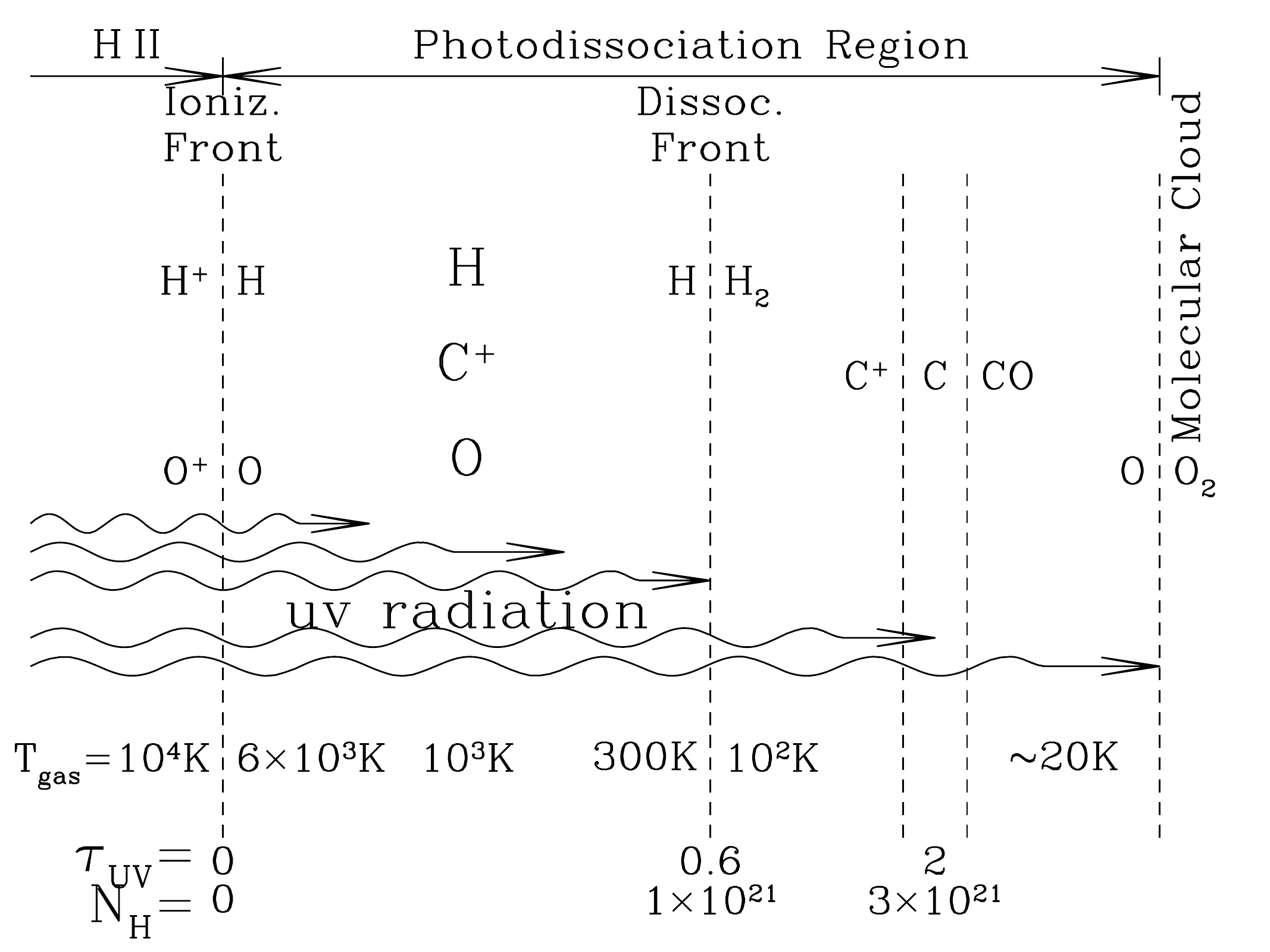}
	\caption{Structure of a photo-dissociation region, the interface between an ionized H\,{\sc ii} region and a molecular cloud. A selection of ionized, atomic and molecular species are shown, together with the gas temperature. Figure from \cite{Draine2011}. 
        } 
	\label{fig_draine}
\end{figure}

In the interstellar medium, stellar radiation from massive stars and cosmic rays can ionize the neutral and diffuse molecular phases, and ultraviolet radiation from massive stars can also dissociate molecules such as $\rm H_2$. 
However, when molecules like $\rm H_2$ in the outer layer of an interstellar cloud absorb photons, they can become optically thick, thereby shielding molecules deeper within the cloud from ionizing stellar light. The interface between the ionized H\,{\sc ii} region surrounding the cloud and the dense molecular cloud is called a {\it photodissociation region}. Its outer boundary is an ionization front, namely the surface where hydrogen is 50\% ionized, and it contains a photodissociation front, where hydrogen is 50\% atomic and 50\% molecular, as shown in Fig.~\ref{fig_draine} from \cite{Draine2011}. 
Dust plays an important role in these regions, since it catalyses $\rm H_2$ formation and can also absorb ionizing ultra-violet photons, thus further protecting molecules against photodissociation. Specific numerical codes, such as the Meudon PDR code \citep{Lepetit2006}, focus on photodissociation regions to calculate their chemical and thermal structure and predict their emission and absorption. In these regions, the different wavelength ranges of the electromagnetic spectrum interact out-of-equilibrium with hundreds of different chemical elements, themselves coupled by thousands of chemical reactions, making such models extremely complex despite the limited size of the region at stake.

\section{Cosmic star formation history}
\label{sect_sfh}

\begin{figure}
	\centering 
        \includegraphics[width=0.48\textwidth, trim={0cm 1.5cm 0cm 1.5cm},clip]{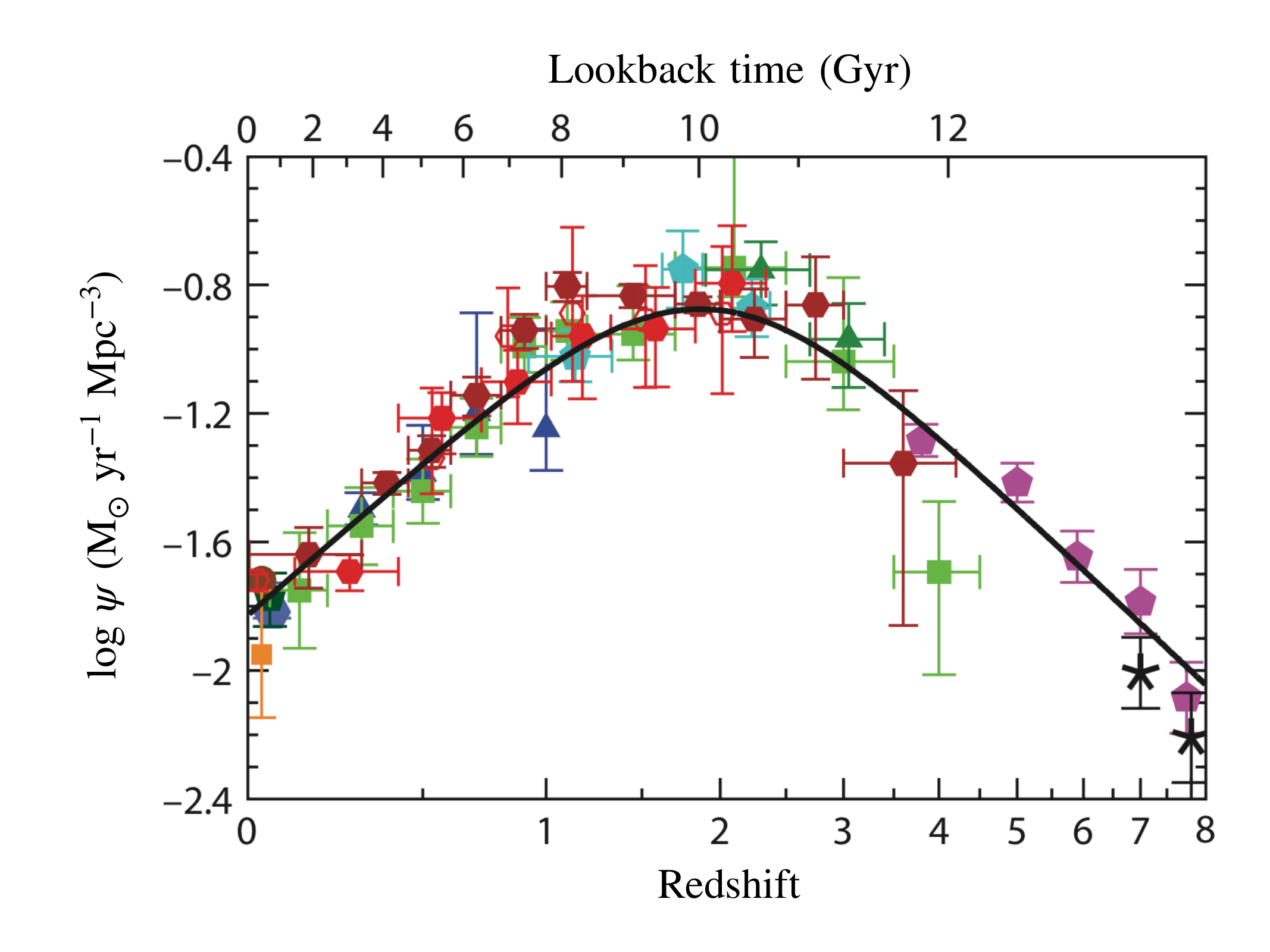}
	\caption{The observed cosmic star formation rate density, i.e. the star formation rate averaged over a comoving volume of the Universe, displays a peak ten billion years ago, around redshift $z\sim 2$. Figure from \cite{Madau2014}. 
        } 
	\label{fig_Madau}
\end{figure}

The current star formation (SFR) of the Milky Way is estimated to be around $2~\rm M_\odot~yr^{-1}$ \citep{Chomiuk2011, Licquia2015}, on par with other nearby spiral galaxies. But galaxies did not always form their stars at similar rates throughout the history of the Universe. Eight billion years ago, typical star-forming galaxies used to form stars at rates around $50~\rm M_\odot~yr^{-1}$; ten billion years ago, those rates were as high as $150~\rm M_\odot~yr^{-1}$ \citep[e.g., ][]{Whitaker2012, Speagle2014}. As shown in Fig.~\ref{fig_Madau}, from \cite{Madau2014}, observations indicate a peak of the star formation activity ten billion years ago, followed by a drop of the star formation rate by an order of magnitude. The high star formation rates observed ten billion years ago imply significant amounts of cold gas available to form stars at the center of dark matter haloes. 
Cosmological simulations suggest that streams stemming from the cosmic web could penetrate deeply into halos of mass lower than a certain threshold and hence bring fresh gas directly at their centers  \citep{Ocvirk2008, Keres2009, Dekel2009}. 
However, direct observational evidence is challenging. These cold streams should be best detectable through the Lyman-$\alpha$ emission of neutral hydrogen atoms \citep{Dijkstra2009}, but deep Lyman-$\alpha$ observations are still scarce and only provide plausible detections of cold streams so far \citep[e.g., ][]{Daddi2021}. 

\begin{figure}
	\centering 
        \includegraphics[width=0.48\textwidth, trim={0.8cm 0.3cm 1.7cm 2.cm},clip]{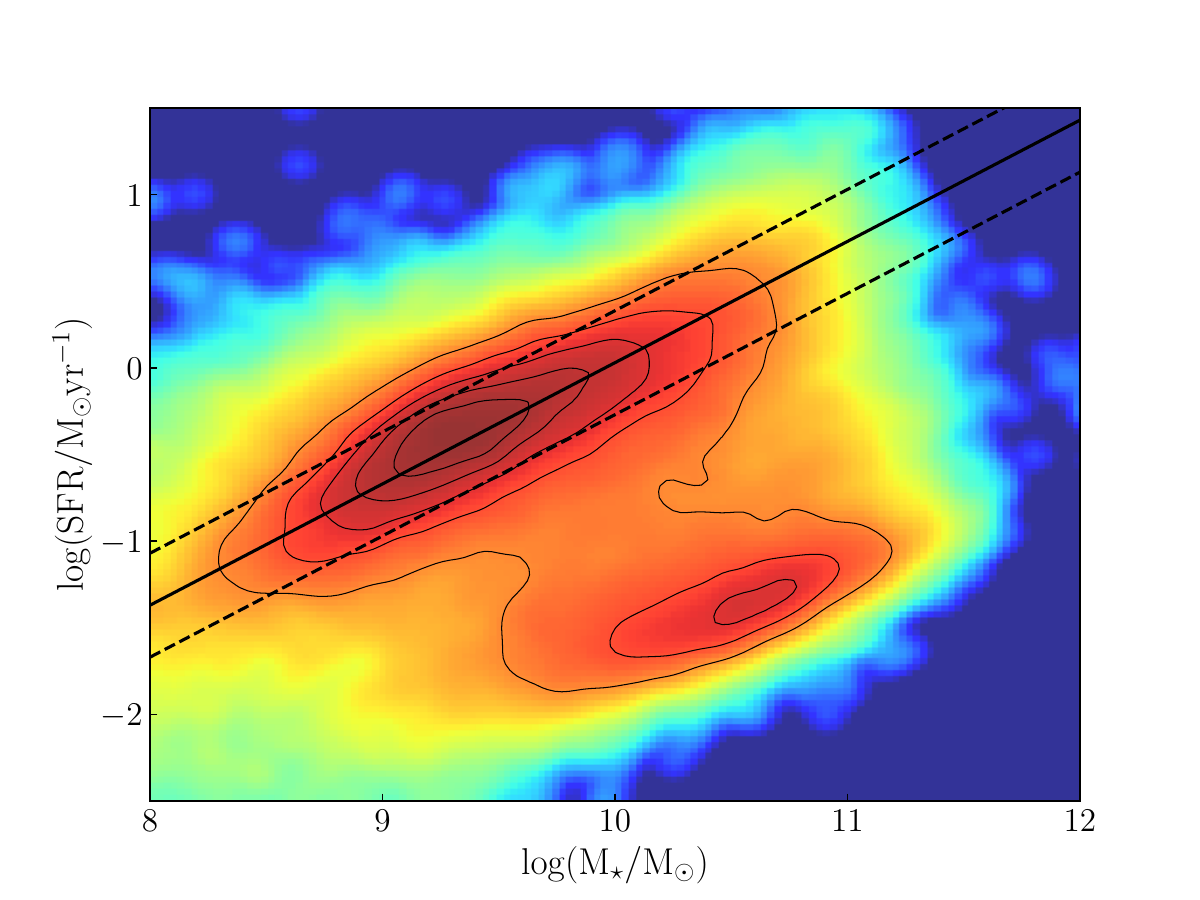}
	\caption{The bimodal distribution of galaxies in the star formation (SFR) versus stellar mass ($M_\star$) plane at $z\sim 0$, using the SDSS       
        \citep{Abazajian2009} MPA-JHU compilation. The color scale and the contours indicate the number of galaxies in each bin, logarithmically. The main sequence of typical star-forming galaxies is in the upper part, the red sequence of galaxies that are forming much less stars at the bottom. The solid line highlights the mean main sequence line at $z=0$ from \cite{Speagle2014}; the dashed lines to the 0.3 dex scatter around this line. 
        } 
	\label{fig_MS}
\end{figure}

Fig.~\ref{fig_Madau} averages the star formation rate over the whole population of galaxies at each epoch. But the galaxy distribution is far from homogeneous, notably with a bimodality between blue star-forming galaxies on one side, the so-called star-formation {\it main sequence} \citep{Brinchmann2004, Noeske2007, Elbaz2007,Whitaker2014}, and red galaxies that form much less stars on the other. The color of a galaxy indeed relates to its stellar population and star formation rate, since young stars are bluer than old ones. Fig.~\ref{fig_MS} shows this bimodality at $z\sim 0$, but it holds at least up to $z\sim 2.5$. 
About 90\% of the cosmic star formation since $z\sim 2.5$, i.e. over the past ten billion years, took place on and around the main sequence \citep{Rodighiero2011, Sargent2012}. In accordance with the cosmic star formation history shown in Fig.~\ref{fig_Madau}, the star formation rate on the main sequence drops by an order of magnitude from $z\sim 2.5$ to present time at any given mass. 
The relative tightness of the sequence promotes an overall smooth and continuous star formation, for instance sustained by accretion along cold streams from the cosmic web and minor mergers. 
Blue main sequence galaxies are predominantly disky while red galaxies are predominantly elliptical \citep[e.g., ][]{Wuyts2011, Wisnioski2015}, which further supports the cold streams scenario since streams would keep rotating disks intact, contrarily to major mergers. 
The low number of galaxies between the two sequences advocates for relatively fast mechanisms that would quench star formation at some point in the history of a galaxy. 
Galaxies above the main sequence, i.e. with particularly high star formation rates, are often associated to mergers, which induce nuclear inflows, shocks, and tidal compression and hence an increase in the star formation rate. However, they may not constitute a distinct population from the main sequence, since mergers do not necessarily result in an offset from the main sequence \citep{Gomez-Guijarro2022, Renaud2022}. Indeed, for a given instantaneous star formation rate, a galaxy which formed more stars in the past may fall within the scatter of the main sequence while a galaxy which formed less stars would appear above without fundamental differences between the two besides their stellar mass.  

\begin{figure}
	\centering 
        \includegraphics[width=0.48\textwidth, trim={0cm 0cm 0cm 0cm},clip]{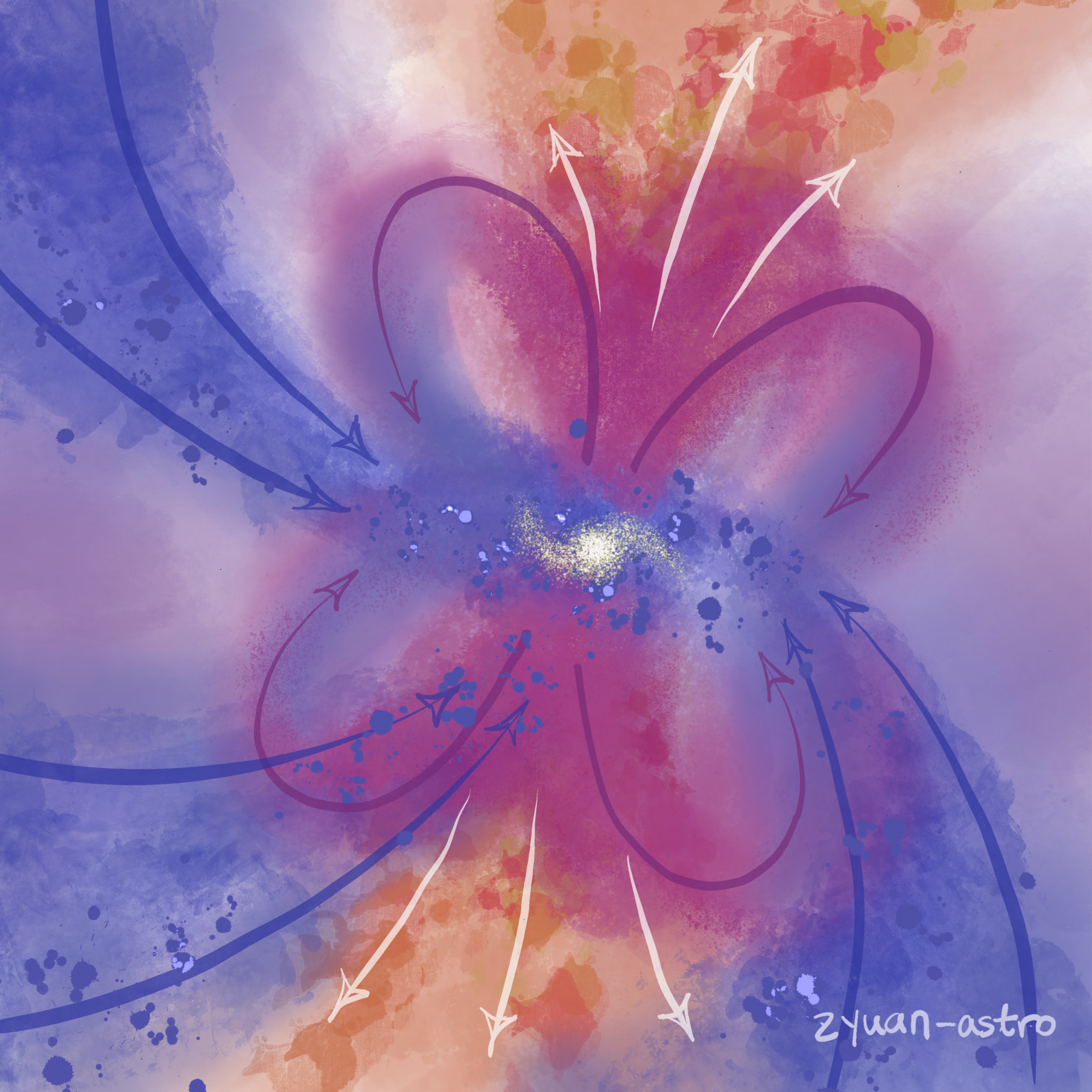}
	\caption{
            Schematic representations of gas flows in and around a galaxy: the galaxy is fed by cold gas streams stemming from the cosmic web (blue arrows), which condense to form stars, while stars and the possible presence of an active galactic nucleus generate powerful outflows that not only hinder subsequent star formation but can also affect the galaxy as a whole and its surrounding circumgalactic medium (yellow/red). Some of the outflowing gas can later fall back towards the galaxy and be recycled (purple). Credits: Zhen Yuan (\href{https://www.zyuan-astro.com/}{zyuan-astro}), inspired by \cite{Tumlinson2017}. 
        } 
	\label{fig_tumlinson}
\end{figure}

Fig.~\ref{fig_tumlinson} illustrates gas cycles within galaxies, with gas channelled to the galaxy along cold streams stemming from the cosmic web, star formation, feedback-driven outflows, and some of the outflowing gas being recycled to form new stars as it falls back on the galaxy.  
As their stellar mass increases, typical galaxies are thought to evolve along the main sequence in a slowly evolving gas-regulated quasi-equilibrium between inflows, outflows, and star formation \citep{Bouche2010, Dave2012, Lilly2013}. Assuming that the mass outflow rate is proportional to the star formation rate to account for stellar feedback processes, the variation in gas mass can be expressed as 
\begin{equation}
\label{eq_bathtub}
\dot{M}_{\rm gas }
=
\dot{M}_{\rm gas,in }
-
{\dot{M}_\star}
-\eta \dot{M}_\star
+ R \dot{M}_\star, 
\end{equation}
where $\dot{M}_{\rm gas,in }$ is the inflow rate from cosmological accretion, ${\dot{M}_\star}$ the star formation rate, $\eta \dot{M}_\star$ the outflow rate, and $R \dot{M}_\star$ the recycling rate. 
Namely, the gas reservoir is fed by accretion and recycling, consumed by star formation, and removed by outflows. Provided that the timescale associated to star formation is smaller than the accretion timescale, this model leads to a steady state solution where the gas mass is constant and the star formation rate proportional to the cosmological inflow rate. Despite its simplicity, it yields a star formation peak around $z\sim 2$, as observed, and approximately the right scaling between star formation rate and stellar mass for the main sequence \citep{Bouche2010}. 

The quasi-equilibrium is expected to last until galaxies enter a denser environment or reach a typical stellar mass $M_\star \sim 10^{11}~M_\odot$, at which point star formation quenches \citep{Peng2010}. In addition to feedback processes from stars and active galactic nuclei, a reduction in gas accretion onto massive haloes, changes in morphology, and environmental effects can contribute to the quenching of star formation.
Cold streams from the cosmic web are notably expected to be unstable above a halo mass of the order of $10^{12}\rm ~M_\odot$ at $z\lesssim 2$, which should shut off the gas supply and prevent further star formation \citep{Dekel2006, Birnboim2007, Dekel2009}. 
As far as morphology is concerned, the build-up of a massive bulge at the center of a galaxy can stabilize the disk against gravitational instabilities and hence quench star formation \citep[{\it morphological quenching};][]{Martig2009}. 
Stellar bars can also influence star formation, although in ways that are still debated: on one side, stirring turbulence could inhibit star formation \citep{Masters2010, Gavazzi2015, Haywood2016, Khoperskov2018, Rosas-Guevara2020}; on the other, driving gas towards the galactic center could enhance star formation there while depleting other regions of gas \citep{Casasola2011, Combes2013, Spinoso2017, George2019, George2020}. 
In dense environments, different physical processes can affect the gas content and star formation: galaxy-galaxy interactions and mergers \citep{Toomre1977}, 
tidal interactions with nearby companions or with the gravitational potential \citep[{\it galaxy harassment}; ][]{Moore1996}, 
ram-pressure \citep{Gunn1972} and tidal stripping of the gas \citep{Byrd1990}, 
a halt in the cold gas supply \citep[{\it strangulation} ; ][]{Larson1980, Peng2015}, 
as well as thermal evaporation due to the interaction between the hot intergalactic gas and the cooler galactic gas \citep{Cowie1977}. 
While a sudden gas removal by ram-pressure or tidal stripping rapidly quenches star formation, the shut-off of gas accretion leads to a slower mode of quenching, since star formation can continue with the gas still available in the galaxy until it is used up \citep{Peng2015}. Tidal interactions and ram-pressure can also locally enhance star formation \citep[e.g.][]{Webb2015, Vulcani2018, Roman-Oliveira2019, Durret2021}. 
Before the final quenching occurs and a galaxy falls to the red sequence, cosmological simulations further suggest star forming galaxies may oscillate within the scatter of the main sequence through episodes of gas compression and enhanced star formation, possibly triggered by major and minor mergers, followed by gas depletion, limited quenching, and replenishment by external accretion \citep{Zolotov2015, Tacchella2016b, Tacchella2016a, Lapiner2023}. 

\section{Efficiency of star formation processes} 
\label{sect_efficiency}

The star formation rate is an {\it extensive} quantity that is expected to correlate with the gas mass of a galaxy, and particularly its molecular gas mass, since stars form through gravitational collapse within giant molecular clouds. To assess star formation processes regardless of the available gas reservoirs, it could be more insightful to use an {\it intensive} quantity reflecting the efficiency of these processes.  
The {\it Kennicutt-Schmidt relation} parametrizes the correlation between the star formation rate and the gas surface densities through a power-law 
\begin{equation}
\Sigma_{\rm SFR} \propto \left(\Sigma_{\rm gas}\right)^N
\end{equation}
\citep{Schmidt1959,Kennicutt1998a,Leroy2008, Kennicutt2012}, where the surface densities can be averaged either over whole galaxies or over resolved sub-regions of the galaxies. One would expect the most fundamental relation to be between the star formation rate and gas {\it volume} densities, as initially proposed by \citet{Schmidt1959}, but observations of external galaxies can only measure {\it surface} densities integrated along the line of sight. 
Observations in local star-forming galaxies indicate a power-law exponent $N=1.41\pm 0.07$ when considering the total atomic and molecular gas content \citep{Kennicutt1998a, delosReyes2019}, and $N=1.0\pm 0.2$ when considering only the molecular gas content \citep{Wong2002, Bigiel2008, Bigiel2011, Leroy2013,Sun2023}. 
A linear relation between the star formation rate and the molecular gas content can be reframed as a constant {\it depletion time}
\begin{equation}
    t_{\rm depl} = \frac{M_{\rm gas}}{{\rm SFR}}, 
\end{equation}
which is the characteristic time needed for a gas mass $M_{\rm gas}$ to turn into stars at constant star formation rate SFR. The depletion time is an intensive measure of how fast star formation proceeds in a given galaxy or sub-region of a galaxy, its inverse sometimes referred to in observational contexts as the {\it star formation efficiency} (${\rm SFE} = 1/t_{\rm depl}$). The mean molecular gas depletion time in the local Universe is around $1-3$ Gyr \citep{Bigiel2008, Leroy2008, Saintonge2011, Saintonge2011b,Sun2023}. 
Under reasonable assumptions, e.g. vertical hydrostatic equilibrium, it may be possible to infer volumic densities and obtain a relation between star formation rate and gas volume densities that is tighter than that between their surface densities \citep{Bacchini2019a, Bacchini2020}, indicating that the relation between volume densities may indeed be more fundamental than the surface but observable Kennicutt-Schmidt relation. 
The relation also breaks down below a few hundreds of parsecs, where star-forming regions and molecular clouds become decorrelated as feedback processes from newly-formed stars rapidly disperse the gas surrounding them (\citealp{Schruba2010, Kruijssen2019, Chevance2020, Kim2022}; see also Fig.~\ref{fig_orion}).

Different theoretical arguments can justify the observed values for the power-law exponent of the Kennicutt-Schmidt relation. The gravitational collapse of a spherical gas cloud of initial volume density $\rho_{\rm gas}$ is expected to occur in a free-fall time $t_{\rm ff} \propto 1/\sqrt{G\rho_{\rm gas}}$, where $G$ is the gravitational constant. Hence, if the star formation rate is a fraction of the star-forming gas mass per free-fall time, one could write its volume density
\begin{equation}
\label{eq_eff}
    \rho_{\rm SFR} = \epsilon_{\rm ff}~ \frac{\rho_{\rm gas}}{t_{\rm ff}}
\end{equation}
where $\epsilon_{\rm ff}$ is a {\it dimensionless star formation efficiency}, preferred in theoretical contexts, corresponding to the star formation rate per free-fall time for a unit gas mass \citep{Krumholz2005, Federrath2012}. 
Comparing the mass of giant molecular clouds with the star formation rate in the Milky Way yields values for $\epsilon_{\rm ff}\sim 1\%$ \citep{Zuckerman1974, Krumholz2007}, which enables to grasp how inefficient star formation is: no more than $\sim 1\%$ of the available gas in giant molecular clouds is transformed into stars per free-fall time. Turbulence, magnetic fields, and feedback processes in the interstellar medium contribute to hinder star formation and reach such a low efficiency compared to purely gravitational free-fall \citep{Federrath2015}. 
Given the expression of the free-fall time, Eq.~(\ref{eq_eff}) implies an exponent 1.5 between the volume densities. Further assuming a uniform line-of-sight scale height leads to an exponent $N=1.5$ between the surface densities, close to the observed Kennicutt-Schmidt relation for the total gas content. 
Instead assuming that all giant molecular clouds are similar and that the star formation rate simply reflects their number leads to a proportionality between star formation rate and gas mass, i.e. $N=1$. 
Other theoretical models enable to physically motivate values of $N$ in the range $0.75-2$ \citep[cf.][]{Bigiel2008}, and more general relations involving for instance the stellar surface density have also been proposed \citep[e.g.][]{Shi2011, Shi2018}. Already-formed stars indeed contribute both to the gravitational potential in which stars form and to the hydrostatic pressure acting on star-forming regions.

\begin{figure}[t]
	\centering 
        \includegraphics[width=0.48\textwidth, trim={0.4cm 0.1cm 3cm 1.2cm},clip]{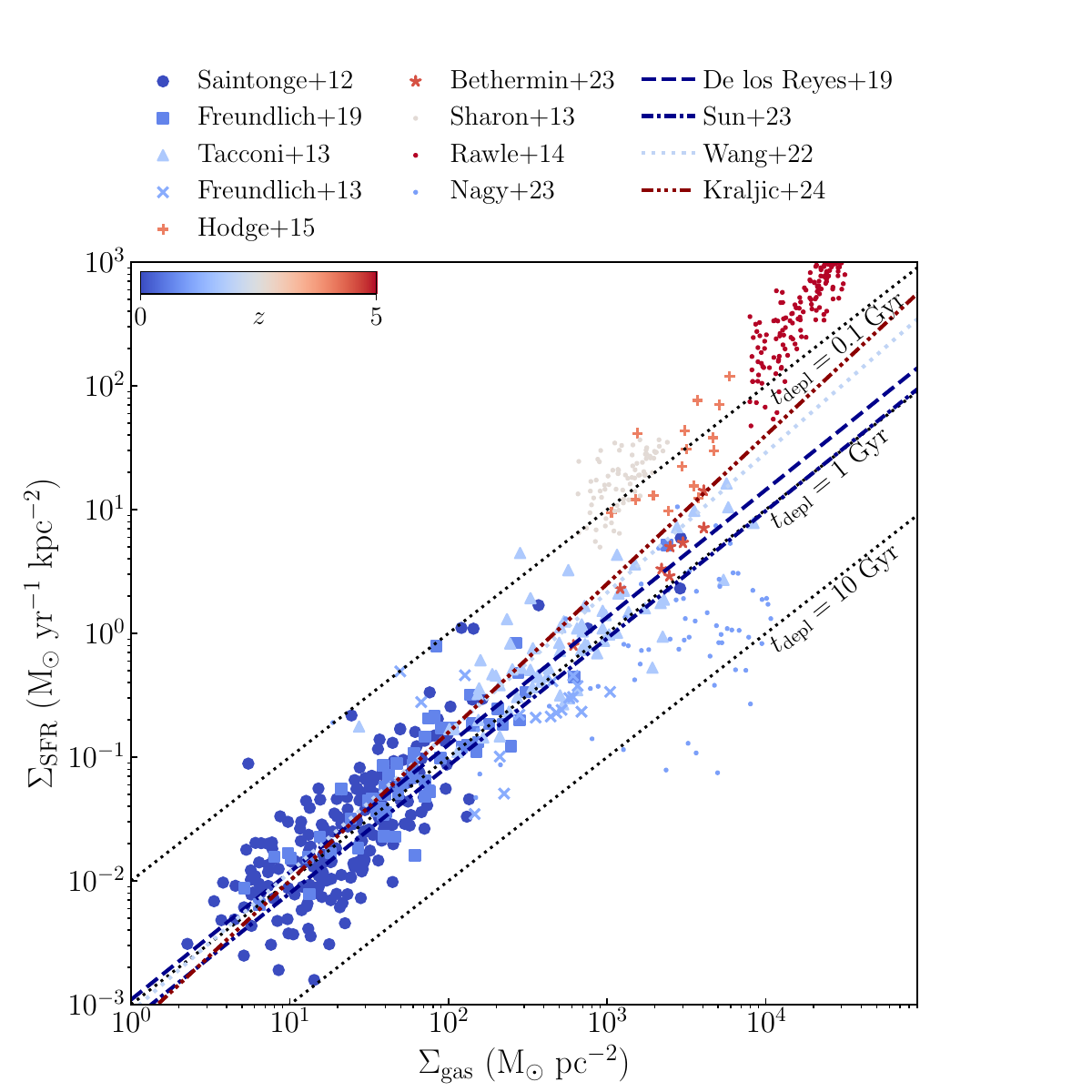}
	\caption{
            Observed Kennicutt-Schmidt relation between the star formation rate and molecular gas surface densities at different epochs. The plot includes galaxy-averaged measurements at $z=0$ \citep{Saintonge2012}, $z=0.5-0.8$ \citep{Freundlich2019}, $z=1.2-2.2$ \citep{Tacconi2013} together with resolved unlensed measurements at $z=1.2$ \citep[][probing ensembles of clumps on scales $\sim$ 8 kpc]{Freundlich2013}, $z=4$ \citep[][on scales $\sim 2$ kpc]{Hodge2015}, and $z=4.5$ \citep[][on $\sim$2 kpc]{Bethermin2023} as well as resolved gravitationally-lensed measurements at $z= 1$ \citep[][down to $\sim$ 100 pc]{Nagy2023}, $z=2.6$ \citep[][on $\sim$6 kpc]{Sharon2013}, and $z=5.2$ \citep[][on $\sim$9 kpc]{Rawle2014}. The color of the data points indicates the average redshift of each sample. The galaxy-averaged \citep{delosReyes2019} and resolved \citep[][on $\sim$1.5 kpc]{Sun2023} Kennicutt-Schmidt relations measured in the local Universe as well as the global relation at $z\sim 4$ from cosmological simulations \citep{Kraljic2024} are shown as dashed and dashed-dotted lines. The diagonal dotted lines indicate constant depletion times $t_{\rm depl}= $ 0.1, 1, and 10 Gyr. 
         } 
	\label{fig_KS}
\end{figure}

As mentioned earlier, the evolution of galaxies throughout the history of the Universe is marked by a peak in star formation activity ten billion years ago. Is the subsequent decline in the star formation rate solely due to a decrease in the amount of gas available to form stars, or is it partially due to less efficient formation processes? 
This can be answered by probing molecular gas at different epochs and determining the associated depletion times. 
Fig.~\ref{fig_KS} shows the Kennicutt-Schmidt relation between the star formation rate and molecular gas surface densities for a selection of samples at different redshifts. The figure includes galaxy-averaged measurements \citep{Saintonge2012, Tacconi2013, Freundlich2019, delosReyes2019}, resolved measurements \citep{Freundlich2013, Hodge2015, Sun2023, Bethermin2023}, resolved measurements in gravitationally-lensed objects \citep{Sharon2013, Rawle2014, Nagy2023}, as well as stacked measurements \citep{Wang2022}. For comparison, it also shows the relation obtained in cosmological simulations by \cite{Kraljic2024}. 
While \cite{Saintonge2012}, \cite{Tacconi2013}, and \cite{Freundlich2019} sample the main sequence relatively evenly by construction, measurements at redshifts higher than 2.5 may however be biased towards outliers. At $z>2.5$, while \cite{Sharon2013}, \cite{Rawle2014}, and \cite{Hodge2015} find relatively low depletion time above the lower-$z$ relation, the measurements of \cite{Bethermin2023} at $z=4.5$ fall precisely on that relation despite galaxies being very different between the two epochs. Gravitationally-lensed measurements possibly suffer from their own lens modeling uncertainties and biases. 
More generally, systematic uncertainties affect star formation rate, stellar and molecular gas mass estimates. 
Nevertheless, the Kennicutt-Schmidt relation seems to be strikingly linear and universal, at least between the peak epoch of star formation and now. 
This indicates that the evolution of the star formation rate throughout the history of the Universe is mainly due to the amount of available molecular gas and that even though this evolution is accompanied by a small decrease in their efficiency, star formation processes are very similar across cosmic time \citep[cf. also][]{Genzel2015, Tacconi2018}.

\section{Conclusion}
\label{sect_conclusion }

Despite large observational programs targeting the molecular gas from which stars form, significant questions remain regarding the role of this gas in the history of star formation and galaxy evolution.
In particular, the reason why galaxies stop forming stars at some point in their life and the transition from the main sequence of star formation to the red sequence are still debated. Does star formation cease because galaxies run out of gas or because the efficiency of star formation drops? 
What is the actual strength of the feedback processes associated with stellar evolution and active galactic nuclei? 
Galaxies are complex systems to simulate, where gravity, turbulence, radiation, star formation and feedback interact with one another, and where the different components and scales affect one another.  State-of-the-art cosmological simulations still fail to harbor gas reservoirs as high as observed and to form stars at the right time and place \citep[e.g.][]{Fensch2021}. Furthermore, they do not form bars as would be expected from observations \citep[e.g.][]{Reddish2022}. 
The fate of star-forming regions is also unclear: are they transient or not, and what is their fate in the galactic disk? If they are long-lived, these structures could notably migrate through dynamical friction towards the center and contribute to the formation of a bulge during the ten billion years following the peak of star formation \citep[e.g.][]{Dekel2009b}.
Finally, not only does the majority of existing observations of the gas concern isolated galaxies that are still actively forming stars, but most observations beyond the local Universe do not distinguish the substructures where stars are forming.

New or upcoming instruments such as the James Webb Space Telescope (JWST) and the Square Kilometer Array (SKA) will enable to probe galaxies and their gas earlier in the history of the Universe and to have a more complete picture of gas in galaxies. JWST has already uncovered galaxies at $z > 10$ \citep[e.g.][]{Finkelstein2022}. ALMA not only enables to spectroscopically confirm the distance of galaxies up to $z\sim 9$ and to detect distant massive dusty galaxies, but also to study gas and star formation processes within galaxies at $z > 4$ on scales smaller than a kiloparsec. Synergies between the two instruments should allow us to observe the distribution of gas and dust together with the stellar populations within distant galaxies and study how these different components are interconnected. SKA will further allow not only the observation of molecular gas at very high redshift but also probing the atomic gas up to the peak epoch of star formation, providing a much more complete picture of gas in galaxies. SKA will indeed detect the HI atomic gas before it cools down to the molecular phase up to $z \sim 1.7$, map environmental effects up to $z\sim 1$, and further allow to better understand the evolution of low surface brightness galaxies \citep{SKA2017}.

\section*{Acknowledgements}
J.F. thanks Zhen Yuan for Fig.~\ref{fig_tumlinson}, Françoise Combes and Florent Renaud for their proofreading and comments, Matthieu Bethermin for part of the code and data underlying Fig.~\ref{fig_KS}, Pierre Ocvirk for Fig.~\ref{fig_reionisation}, and the anonymous referees for their careful reading and constructive comments. 




\bibliographystyle{elsarticle-harv} 
\bibliography{Freundlich_FPP2023}






\end{document}